\UseRawInputEncoding
\documentclass[UTF8,aps,prl,groupedaddress,twocolumn,showpacs]{revtex4-1}
\usepackage[dvips]{graphicx}
\usepackage{dcolumn}
\usepackage{bm}
\usepackage{amssymb}
\usepackage{epstopdf}
\usepackage{amsmath}
\usepackage{color}

\begin{document}
\title{A Scheme to fabricate magnetic graphene-like cobalt nitride CoN$_4$ monolayer proposed by first-principles calculations}
\author{Ming Yan$^{1}$}
\author{Shuo Zhang$^{2}$}
\author{Yanyun Wang$^{2}$}\email{sxx430@126.com}
\author{Fengjie Ma$^{3}$}\email{fengjie.ma@bnu.edu.cn}
\affiliation{$^{1}$Department of Mechanical and Aerospace Engineering, Syracuse University, Syracuse, NY 13244, USA}
\affiliation{$^{2}$College of Physics and Engineering, Qufu Normal University, Qufu, Shandong 273165, China}
\affiliation{$^{3}$The Center for Advanced Quantum Studies and Department of Physics, Beijing Normal University, Beijing 100875, China}
\begin{abstract}
We propose a scheme to fabricate the cobalt nitride CoN$_4$ monolayer, a magnetic graphene-like two-dimensional material,
in which all Co and N atoms are in a plane. Under the pressure above 40 GPa, the bulk CoN$_4$ is stabilized in a triclinic phase. With the pressure decreasing, the triclinic phase of CoN$_4$ is transformed into an orthorhombic phase, and the latter is a layered compound with large interlayer spacing. At ambient condition, the weak interlayer couplings are so small that single CoN$_4$ layer can be exfoliated by the mechanical method.

\end{abstract}


\maketitle
In 2017, two-dimensional (2D) ferromagnetism was first realized experimentally in the CrI$_3$ monolayer[\cite{Huang2017}], which is a milestone in the field of 2D materials. CrI$_3$ monolayer is composed of three layers of atoms and it is not the thinnest material. Therefore, to fabricate magnetic 2D materials with the thinness limit is an important issue.

Graphene and the graphene-like 2D materials with a single-atom-thickness,
such as borophene, boron nitride, and $g$-C$_3$N$_4$，are extremely promising for next-generation electronic devices[\cite{Mannix2015,Song2010,Groenewolt2005}]. But there is no intrinsic magnetism with long-range orders in them because they are made up of the third to fifth group elements,
which greatly limits the application range of these graphene-like materials.
The most direct idea is to place the magnetic atoms in the sheet of graphene-like materials.
When a transition metal atom is inserted in the graphene sheet or substitutes a carbon atom of graphene, it is difficult to reside in the same plane [\cite{Krasheninnikov2009}]. The main reason is that the metal atom can not be bonded to the C atoms with a planar pattern.
However, if we adopt nitrogen, rather than carbon, transition metal atom may be bonded to nitrogen to form a planar monolayer.
The idea is supported by the experiments of the single-atom catalyst synthesis, in which the planar geometry of the $M$N$_4$ ($M$ = Fe, Co, Ni, etc.) moieties implanted into porous graphene sheets has been identified by the annular dark-field scanning transmission electron microscope [\cite{Fei2018}].
There are also some theoretical studies to provide the evidences for the stability of the planar $M$N$_4$ ($M$ = Cr, Mn, Co, Pt, etc.) moieties.
For example, the CoN$_4$C$_{10}$, CoN$_4$C$_2$, CrN$_4$C$_2$, Mn$_2$N$_6$C$_6$, Pt$_2$N$_8$C$_6$ monolayers containing the $M$N$_4$ moieties [\cite{Liu2021,Liu2021a,Liu2021b,Feng2022,Dong2022}] have been proved to be thermally, dynamically, and mechanically stable by the calculations of formation energy, molecular dynamics, phonon, and elastic constants.
Moreover, the binary transition metal nitrides MN$_4$ ($M$ = V, Cr, Mn, Fe, or Co) monolayers, which possess the similar structure to the graphene or graphene allotrope but contain the transition metal elements, have been predicted in the recent theoretical studies [\cite{Liu2022,Zhang2022}].
Despite the great progress in this field, how to fabricate graphene-like magnetic two-dimensional monolayers experimentally is a crucial problem at this stage.

Very recently, the triclinic phase of beryllium tetranitride BeN$_4$ was synthesized at the pressure of 85 GPa [\cite{Bykov2021}], and upon decompression to ambient condition, it transforms into a layered compound with the weak van der Waals interactions.
A single BeN$_4$ layer made up of beryllium atoms and polymeric nitrogen chains is proposed by the experiments and is confirmed by the first-principles calculations [\cite{Bykov2021,Mortazavi2021}].
The experiment of BeN$_4$ synthesis motivates us to think about whether the transition metal nitride $M$N$_4$ monolayers can be fabricated according to the above scheme.
In experiments, FeN$_4$ under high pressure crystallizes into the triclinic structure with $P \bar 1$ group symmetry [\cite{Bykov2018,Bykov2018a,Jiao2020}]. In the theoretical studies, the CoN$_4$, MgN$_4$, and ZnN$_4$ structures under high pressure are predicted to be $P \bar 1$ triclinic phase through systematic structure search in CALYPSO package  [\cite{Liu2021x,Wei2017,Shi2020}].
According to the above studies, the structures of FeN$_4$, CoN$_4$, MgN$_4$, and ZnN$_4$ at high pressure are definitely determined, which is a necessary prerequisite for our fabrication scheme of magnetic monolayer.
Because the structures of FeN$_4$ and ZnN$_4$ at ambient pressure are not layered materials with large interlayer spacing and there is no magnetism in MgN$_4$ and ZnN$_4$, we focus on CoN$_4$ compound.

In this work,
on the base of the first-principles calculations, we demonstrate that the graphene-like CoN$_4$ monolayer can be fabricated by the three steps, high-pressure synthesis, decompression to ambient condition, and mechanical exfoliation.

\begin{figure}
\begin{center}
\includegraphics[width=7.50cm]{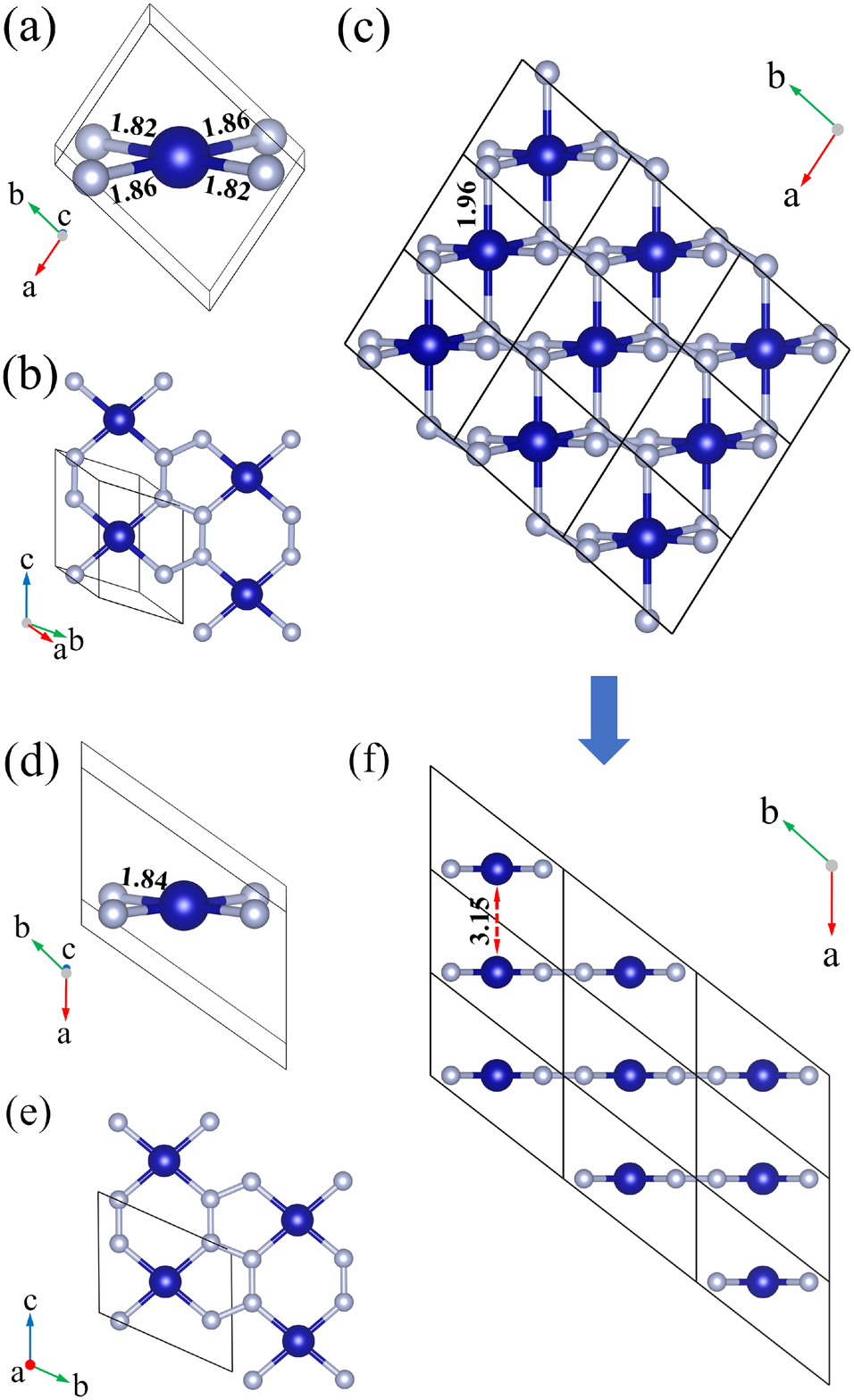}
\caption{Crystal structure of $P \bar 1$-CoN$_4$ at 40 GPa: (a) the unit cell; (b) top view of CoN$_4$ layer; (c) side view along $c$ direction.
Crystal structure of $Cmmm$-CoN$_4$ at 0 GPa: (d) the unit cell; (e) top view of CoN$_4$ layer; (f) side view along $c$ direction.
The numbers are the bond length or interlayer distance and the unit is \AA. } \label{struct12}
\end{center}
\end{figure}

 The electronic structure calculations are performed with the VASP software package, in which the plane wave pseudopotential method and the projector augmented-wave (PAW) pseudopotential with Perdew-Burke-Ernzerhof (PBE) exchange-correlation functional [\cite{PhysRevB.47.558, PhysRevB.54.11169, PhysRevLett.77.3865, PhysRevB.50.17953}] are used.
The plane wave basis cutoff is set to 600 eV, and the convergence thresholds of total energy and force are set to 10$^{-5}$ eV and 0.01 eV/\AA.
The interlayer distance was set to 20 \AA~ and a mesh with the density of 30 k-points per \AA $^{-1}$ is used for the Brillouin zone integration.
The phonon calculations are carried out with the supercell method in the PHONOPY program,
and the real-space force constants were calculated using density-functional perturbation theory (DFPT) as implemented in VASP  [\cite{Togo2015}].
In the ab initio molecular dynamics simulations,
the 2 $\times$ 2 $\times$ 2 supercells were employed and the temperature was kept at 1000 K for 5 ps with a time step of 1 fs in the canonical ensemble (NVT) [\cite{Martyna1992}].

We first study the atomic structure of CoN$_4$ compound under pressure.
The structure of $P \bar 1$-CoN$_4$ is displayed in Fig. \ref{struct12} (a), (b), and (c), and the unit cell is marked by solid lines. One unit cell is composed of only one CoN$_4$ moiety.
 Due to the periodicity of the crystal structure, the CoN$_4$ moieties connect each other along the \{110\} lattice plane to form a infinite CoN$_4$ monolayer.
 The thing to notice is that at the pressure of 40 GPa the CoN$_4$ moiety in each unit cell still remains a planar configuration, which reveals the robustness of the CoN$_4$  planar configuration.
 The small buckling deformation occurs at the junction between two adjacent unit cells, leading to the fluctuation of the CoN$_4$ layer.
 The four N-Co bonds in the CoN$_4$ moiety have the length of 1.82 \AA~ and 1.86 \AA. The interlayer distance is 1.96 \AA~, equal to the distance of Co and N between two adjacent layers, and the distance increases gradually with the pressure decreasing.

\begin{figure}
\begin{center}
\includegraphics[width=8.0cm]{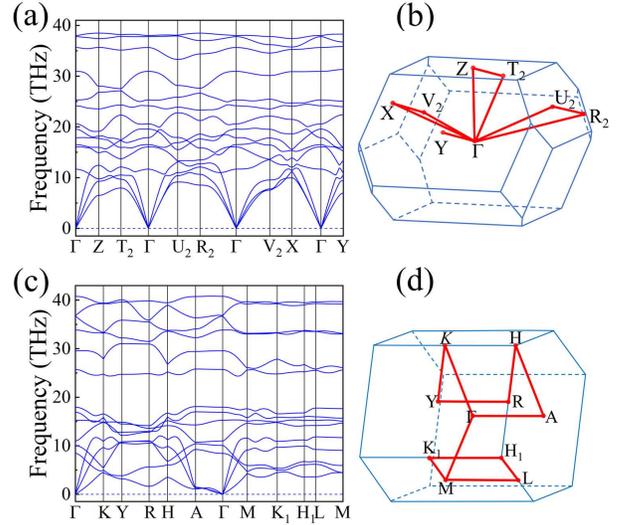}
\caption{(a)-(b) Phonon spectra of $P \bar 1$-CoN$_4$ at 40 GPa and the Brillouin zone.  (c)-(d) Phonon spectra of $Cmmm$-CoN$_4$ at 0 GPa and the Brillouin zone.
 } \label{phonon-press}
\end{center}
\end{figure}

 To examine the dynamic stability of $P \bar 1$-CoN$_4$ structure, we perform the calculations of phonon spectra at 40 GPa.
 The phonon bands are shown in Fig. \ref{phonon-press}(a), and there is no imaginary frequency for their phonon modes, indicating that $P \bar 1$-CoN$_4$ under the pressure from 40 GPa is dynamically stable.
For $P \bar 1$-CoN$_4$ at 40 GPa, the relative formation enthalpy $\Delta H$ with respect to Co metal and N$_2$ are computed in terms of the following formula \ref{Hxxx},
\begin{equation} \label{Hxxx}
\Delta H = (H_{CoN_4} - H_{Co} -2H_{N_2}) / 5.
\end{equation}
where $H_{CoN_4}$, $H_{Co}$, and $H_{N_2}$ are the enthalpies of CoN$_4$, Co metal, and N$_2$ at 40 GPa, respectively.
The $\Delta H$ is -0.45 eV/atom and the negative values indicate that $P \bar 1$-CoN$_4$ structure is energetically favorite at 40 GPa.
The hydrostatic pressure of 40 GPa is easily achieved experimentally. Similar to the conditions for FeN$_4$ and BeN$_4$ synthesis [\cite{Bykov2018, Bykov2018a, Bykov2021}], we can use a BX90 diamond anvil cell (DAC) to create a high pressure environment with N$_2$ as a pressure-transmitting medium. Besides, the multianvil large-volume press (LVP), such as Kawai-type multianvil apparatus, can be used to generate the pressure of 40 GPa in a larger sample chamber[\cite{Bykov2019}].

With the decompression to 0 GPa, the triclinic $P \bar 1$-CoN$_4$ structure will evolve into a orthorhombic structure with the $Cmmm$ group symmetry.
The scenario is different from the BeN$_4$ compound, which still remains the triclinic structure when the pressure decrease from 85 GPa to ambient pressure in experiments [\cite{Bykov2021}].
The $Cmmm$-CoN$_4$ structure is shown in Fig. \ref{struct12}(d), (e), and (f). Its unit cell is composed of only one formula cell and the CoN$_4$ moiety keeps a planar geometry. Compared with the $P \bar 1$-CoN$_4$ structure at 40 GPa, the small buckling deformation at the junction of two CoN$_4$ moieties disappears and the whole CoN$_4$ layer become a perfectly planar configuration.
In addition, the upper layer in the $Cmmm$-CoN$_4$ structure has a obvious translation with respect to the lower layer, causing the Co atoms in the adjacent layers to face to each other.
More importantly, the interlayer distance increases from 1.96 \AA~ to 3.15 \AA.
Next, we verify the stability of $Cmmm$-CoN$_4$ structure under ambient pressure. Fig. \ref{phonon-press}(c) shows the phonon spectra. There is no imaginary frequency in the phonon curves, which proves that the structure is dynamically stable.
 The formation enthalpy $\Delta H$ with respect to Co metal and N$_2$ are computed and the value is 0.40 eV/atom,
 which is comparable to the ones of other nitrides synthesized in experiments, such as CuN$_3$ [\cite{Wilsdorf1948}](0.55 eV/atom), PtN$_2$ [\cite{Crowhurst2006}] (0.42 eV/atom), and $g$-C$_3$N$_4$ [\cite{Groenewolt2005}] (0.35 eV/atom).
 So, the $Cmmm$ structure is a metastable phase similar to the above nitrides, which can exist at ambient condition.

\begin{figure}
\begin{center}
\includegraphics[width=8.0cm]{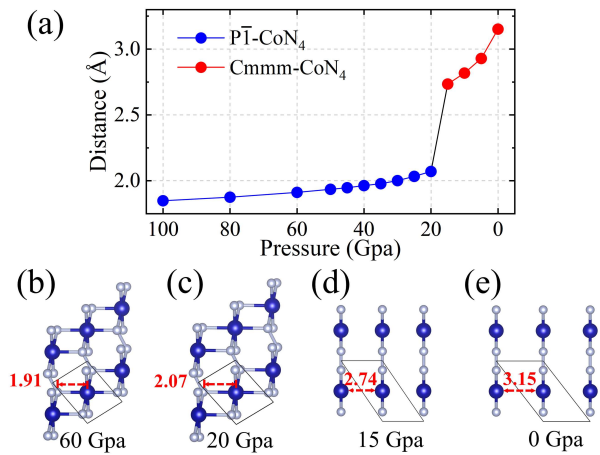}
\caption{(a) Interlayer distance of CoN$_4$ in $P \bar 1$ and $Cmmm$ phases varying with respect to the pressure. (b) - (e), At the selected pressures, the interlayer distances are marked with red numbers. Their unit is \AA.
 } \label{distance}
\end{center}
\end{figure}

The variations of interlayer distance with pressure in $P \bar 1$-CoN$_4$ and $Cmmm$-CoN$_4$ structures are shown in Fig. \ref{distance}(a).
The blue and red solid circles are related to the $P \bar 1$-CoN$_4$ phase and $Cmmm$-CoN$_4$ phase, and the selected interlayer distances are marked in Fig. \ref{distance}(b), (c), (d), and (e).
Above 20 GPa, the CoN$_4$ retains the $P \bar 1$-CoN$_4$ structure, while it shifts to the $Cmmm$-CoN$_4$ structure below 15 GPa.
A sharp increase of interlayer distance occurs at about 15 GPa, and the distance reaches up to 3.15 \AA~ at 0 GPa, which means that there is very weak coupling between adjacent CoN$_4$ layers.
The weak interlayer coupling and large interlayer space are similar to the case of triclinic BeN$_4$ compound under ambient condition.
We can imagine the following process of preparing CoN$_4$ monolayer.
 The $P \bar 1$-CoN$_4$ phase is first synthesized under high pressure, then we gradually decompress the pressure to 0 GPa. The interlayer distance increases slowly and the Co-N bonds in the layer are not destroyed during the decompression process. At last, the $Cmmm$-CoN$_4$ phase with the planar CoN$_4$ layer and large interlayer spacing is obtained.

\begin{figure}
\begin{center}
\includegraphics[width=7.0cm]{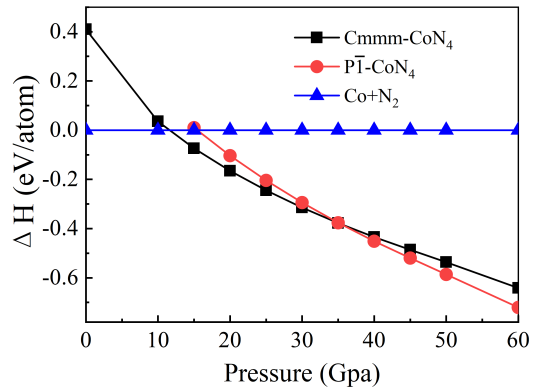}
\caption{ The enthalpies of $P \bar 1$-CoN$_4$ and $Cmmm$-CoN$_4$ phases at different pressures. The enthalpy sum of two components, Co metal and N$_2$, is set to zero.
 } \label{enthalpy}
\end{center}
\end{figure}

On the other hand, we compute and compare the enthalpies of $P \bar 1$-CoN$_4$ and $Cmmm$-CoN$_4$ phases at different pressures, and the data are shown in Fig. \ref{enthalpy}.
When the pressures are larger than about 15 GPa for $P \bar 1$-CoN$_4$ or 12 GPa for $Cmmm$-CoN$_4$ phase, the relative enthalpies are negative, indicating that the synthesis of CoN$_4$ requires a pressure environment.
With the pressure higher than 35 GPa, $P \bar 1$-CoN$_4$ phase has the lower enthalpy than $Cmmm$-CoN$_4$ phase, indicating that $P \bar 1$-CoN$_4$ phase is the dominated structural phase.
When the pressure is less than 35 GPa, the $Cmmm$-CoN$_4$ phase is preferred energetically.
In our calculations of structural optimization at 15 GPa, the $P \bar 1$-CoN$_4$ phase will be automatically transformed into the $Cmmm$-CoN$_4$ phase, which indicates that the $P \bar 1$-CoN$_4$ phase does not exist under low pressure, which is consistent with the previous study [\cite{Liu2021x}].
On the contrary, the $Cmmm$-CoN$_4$ phase is a favorable structure at ambient condition.

Since there exists large interlayer space in CoN$_4$ compound, one natural question is whether single CoN$_4$ layer can be exfoliated from the bulk counterpart.
 Jia \textit{et. al} proposed a method to assess the exfoliability of bulk crystals[\cite{Jia2021}], in which beside cleavage energy, the mechanical integrity during the exfoliation was also taken into account.
 The ratio $\sigma_s/\gamma$ is a measure to judge the exfoliability of layered compounds. $\sigma_s$ is the intrinsic intralayer strength of the material, a parameter to ensure the in-plane mechanical integrity during mechanical exfoliation, and $\gamma$ is the cleavage energy density.
 For simplicity, $\sigma_s/\gamma$ is represented by the in-plane Young's modulus $Y_{in}$ and out-plane modulus $Y_{out}$, i.e. $\sigma_s/\gamma \approx  Y_{in}/Y_{out}$, which makes it easier to judge the exfoliability.
 By computing the values of $Y_{in}/Y_{out}$ for 34 already-exfoliated 2D materials in experiments, the exfoliable threshold is determined to be 10.5.
If the $Y_{in}/Y_{out}$ value is greater than 10.5, the material are regarded to be exfoliable[\cite{Jia2021}].
 For the $Cmmm$-CoN$_4$, we perform the calculations of the elastic constants and the Young's modulus are derived.
The Young's modulus $Y_{in}$ in the plane of CoN$_4$ layer varies from 600.2 to 814.6 GPa and the out-plane modulus $Y_{out}$ is 19.6 GPa.
The $Y_{in}/Y_{out}$ is in the range of 30.2 - 41.6 GPa, greater than the threshold of 10.5.
Therefore, the exfoliability of $Cmmm$-CoN$_4$ compound is confirmed.

The next question is whether the free-standing CoN$_4$ monolayer is stable.
In the previous study [\cite{Zhang2022}], we have carried out the calculations of the phonon spectra and the molecular dynamics simulations for the free-standing CoN$_4$ monolayer. No imaginary frequency emerges in the phonon spectra and the integrity of planar CoN$_4$ layer is not broken after 10 ps at 1000 K, which demonstrates that CoN$_4$ monolayer has good dynamical and thermal stability.
Furthermore, the stability mechanism of the CoN$_4$ monolayer is explained by the cooperation of Co-N coordination bond, $\pi$ conjugation on the armchair N-chain, and $\pi$-d conjugation effect between N-N $\pi$ bond and $d$ orbitals of Co atom.
By now, we have demonstrate that the graphene-like CoN$_4$ layer can be exfoliated from the bulk counterpart and can exist in the form of a single free-standing layer.

\begin{figure}[htbp]
\begin{center}
\includegraphics[width=7.0cm]{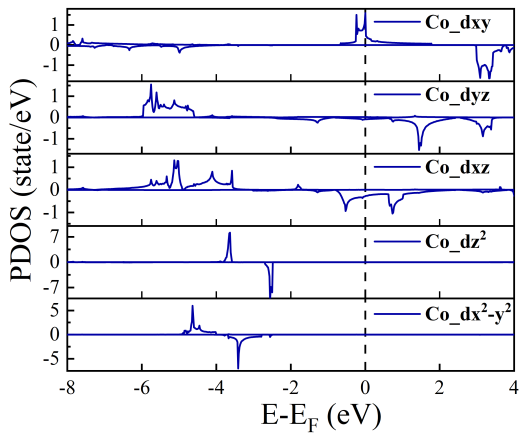}
\caption{Partial density of states of Co 3$d$ suborbitals of one Co atom in CoN$_4$ monolayer. Positive and negative values correspond to the spin-up and spin-down states.
 } \label{pdos}
\end{center}
\end{figure}

\begin{figure}[htbp]
\begin{center}
\includegraphics[width=7.0cm]{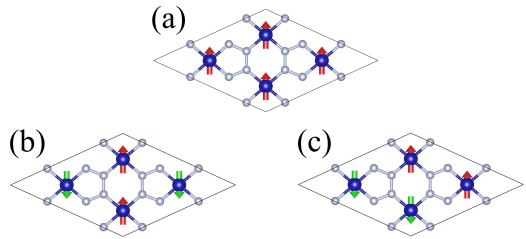}
\caption{Sketches of three magnetic orders in CoN$_4$ monolayer: (a) FM, (b) AFM-I, and (c) AFM-II.
 } \label{orders}
\end{center}
\end{figure}

The GGA + U method is employed to investigate the magnetic properties of CoN$_4$ monolayer. By the self-consistent calculation with linear response method, the value of Hubbard U is determined to be 5.1 eV.
Fig. \ref{pdos} shows the projected density of states on Co five $3d$ orbitals. As can be seen, there is the obvious spin-splitting between spin-up and spin-down states, and the magnetic moment of Co atom are mainly from the $d_{xz}$, $d_{yz}$, and $d_{xy}$ orbitals.
We also compute the energies for ferromagnetic order (FM) and two antiferromagnetic orders (AFM-I and AFM-II), shown in Fig. \ref{orders}. The FM, AFM-I, and AFM-II energies per formula cell are -34.17, -34.80, and -34.69 eV, respectively.
The AFM-I and AFM-II orders have lower energy than FM order, indicating that the magnetism in the ground state is antiferromagnetic.

In summary, based on the first-principles calculations, we propose a fabrication scheme of single-atom-thick magnetic 2D material, graphene-like CoN$_4$ monolayer, which is a transition metal nitride sheet with the antiferromagnetism.
Under the pressure above 35 GPa, the CoN$_4$  adopts a riclinic phase with the $P \bar 1$ group symmetry, while the $Cmmm$ phase is more favorite energetically with the pressure lower than 35 GPa. When the pressure decrease to 0 GPa, the interlayer spacing in the $Cmmm$-CoN$_4$ phase can reach up to 3.15 \AA.
The exfoliation of single CoN$_4$ layer from the bulk counterpart is demonstrated by the ratio $Y_{in}/Y_{out}$.
In addition, the antiferromagnetism is determined by the spin-polarized calculations.
Therefore, our studies concerning the fabrication scheme of the graphene-like CoN$_4$ monolayer with the intrinsic magnetism provide a new way to the synthesis of magnetic 2D materials in experiments.

This work was supported by the National Natural Science Foundation of China under Grants No. 12074040, No. 11974207, and No. 12274255.


\begin{thebibliography}{0}%
\makeatletter
\providecommand \@ifxundefined [1]{%
 \@ifx{#1\undefined}
}%
\providecommand \@ifnum [1]{%
 \ifnum #1\expandafter \@firstoftwo
 \else \expandafter \@secondoftwo
 \fi
}%
\providecommand \@ifx [1]{%
 \ifx #1\expandafter \@firstoftwo
 \else \expandafter \@secondoftwo
 \fi
}%
\providecommand \natexlab [1]{#1}%
\providecommand \enquote  [1]{``#1''}%
\providecommand \bibnamefont  [1]{#1}%
\providecommand \bibfnamefont [1]{#1}%
\providecommand \citenamefont [1]{#1}%
\providecommand \href@noop [0]{\@secondoftwo}%
\providecommand \href [0]{\begingroup \@sanitize@url \@href}%
\providecommand \@href[1]{\@@startlink{#1}\@@href}%
\providecommand \@@href[1]{\endgroup#1\@@endlink}%
\providecommand \@sanitize@url [0]{\catcode `\\12\catcode `\$12\catcode
  `\&12\catcode `\#12\catcode `\^12\catcode `\_12\catcode `\%12\relax}%
\providecommand \@@startlink[1]{}%
\providecommand \@@endlink[0]{}%
\providecommand \url  [0]{\begingroup\@sanitize@url \@url }%
\providecommand \@url [1]{\endgroup\@href {#1}{\urlprefix }}%
\providecommand \urlprefix  [0]{URL }%
\providecommand \Eprint [0]{\href }%
\providecommand \doibase [0]{http://dx.doi.org/}%
\providecommand \selectlanguage [0]{\@gobble}%
\providecommand \bibinfo  [0]{\@secondoftwo}%
\providecommand \bibfield  [0]{\@secondoftwo}%
\providecommand \translation [1]{[#1]}%
\providecommand \BibitemOpen [0]{}%
\providecommand \bibitemStop [0]{}%
\providecommand \bibitemNoStop [0]{.\EOS\space}%
\providecommand \EOS [0]{\spacefactor3000\relax}%
\providecommand \BibitemShut  [1]{\csname bibitem#1\endcsname}%
\let\auto@bib@innerbib\@empty
\end{thebibliography}%


\begin{thebibliography}{99}
\bibitem{Huang2017} B. Huang, G. Clark, E. Navarro-Moratalla, D. R. Klein, R. Cheng, K. L. Seyler, D. Zhong, E. Schmidgall, M. A. McGuire, D. H. Cobden, W. Yao, D. Xiao, P. Jarillo-Herrero, and X. Xu, Nature 546, 270 (2017).
\bibitem{Mannix2015} A. J. Mannix, X.-F.
Zhou, B. Kiraly, J. D. Wood, D. Alducin, B. D. Myers, X. Liu, B. L. Fisher, U. Santiago, J. R. Guest, M. J. Yacaman, A. Ponce, A. R. Oganov, M. C. Hersam, and N. P. Guisinger, Science 350, 1513 (2015)
\bibitem{Song2010} L. Song, L. Ci, H. Lu, P. B. Sorokin, C. Jin, J. Ni, A. G. Kvashnin, D. G. Kvashnin, J. Lou, B. I. Yakob Ajayan, Nano Letters 10, 3209 (2010)
\bibitem{Groenewolt2005} M. Groenewolt and M. Antonietti, Advanced Materials 17, 1789 (2005).



\bibitem{Krasheninnikov2009} A. V. Krasheninnikov, P. O. Lehtinen, A. S. Foster, P. Pyykk?, and R. M. Nieminen, Physical Review Letters 102, 126807 (2009).
8
\bibitem{Fei2018} H. Fei, J. Dong, Y. Feng, C. S. Allen, C. Wan, B. Volosskiy, M. Li, Z. Zhao, Y. Wang, H. Sun, P. An, W. Chen, Z. Guo, C. Lee, D. Chen, I. Shakir, M. Liu, T. Hu, Y. Li, A. I. Kirkland, X. Duan, and Y. Huang, Nature Catalysis 1, 63 (2018).

\bibitem{Liu2021} D. Liu, S. Zhang, M. Gao, and X.-W. Yan, Physical Review B 103, 125407 (2021).
\bibitem{Liu2021a} D. Liu, P. Feng, M. Gao, and X.-W. Yan, Physical Review B 103, 155411 (2021).
\bibitem{Liu2021b} D. Liu, S. Zhang, M. Gao, X.-W. Yan, and Z. Y. Xie, Applied Physics Letters 118, 223104 (2021).
\bibitem{Feng2022} P. Feng, S. Zhang, D. Liu, M. Gao, F. Ma, X.-W. Yan, and Z. Y. Xie, The Journal of Physical Chemistry C 126, 10139 (2022).
\bibitem{Dong2022} J. Dong, C. Wang, X. Zhao, M. Gao, X.-W. Yan, F. Ma, and Z.-Y. Lu, Physical Review Materials 6, 074202 (2022).
\bibitem{Liu2022} D. Liu, P. Feng, S. Zhang, M. Gao, F. Ma, X.-W. Yan, and Z. Y. Xie, Physical Review B 106 125421 (2022).
\bibitem{Zhang2022} S. Zhang, P. Feng, D. Liu, H. Wu, M. Gao, T. Xu, X.-W. Yan, and Z. Y. Xie, (2022), arXiv:2204.13551v1.
\bibitem{Bykov2021} M. Bykov, T. Fedotenko, S. Chariton, D. Laniel, K. Glazyrin, M. Hanfland, J. S. Smith, V. B. Prakapenka, M. F. Mahmood, A. F. Goncharov, A. V. Ponomareva, F. Tasnádi, A. I. Abrikosov, T. Bin Masood, I. Hotz, A. N. Rudenko, M. I. Katsnelson, N. Dubrovinskaia, L. Dubrovinsky, and I. A. Abrikosov, Physical Review Letters 126, 175501 (2021).
\bibitem{Mortazavi2021} B. Mortazavi, F. Shojaei, and X. Zhuang, Materials Today Nano 15, 2 (2021).

\bibitem{Bykov2018}
M. Bykov, E. Bykova, G. Aprilis, K. Glazyrin, E. Koemets, I. Chuvashova, I. Kupenko, C. McCammon, M. Mezouar, V. Prakapenka, H. P. Liermann, F. Tasn′ adi, A. V. Ponomareva, I. A. Abrikosov, N. Dubrovinskaia, and L. Dubrovinsky, Nature Communications 9, 1 (2018).
\bibitem{Bykov2018a}
M. Bykov, S. Khandarkhaeva, T. Fedotenko, P. Sedmak, N. Dubrovinskaia, and L. Dubrovinsky, Acta Crystallographica Section E: Crystallographic Communications 74, 1392 (2018).
\bibitem{Jiao2020}
F. Jiao, C. Zhang, and W. Xie, Journal of Physical Chemistry C 124, 19953 (2020).
\bibitem{Liu2021x}
S. Liu, R. Liu, H. Li, Z. Yao, X. Shi, P. Wang, and B. Liu, Inorganic Chemistry 60, 14022 (2021).
\bibitem{Wei2017}
S. Wei, D. Li, Z. Liu, X. Li, F. Tian, D. Duan, B. Liu, and T. Cui, Physical Chemistry Chemical Physics 19, 9246 (2017).
\bibitem{Shi2020}
X. Shi, Z. Yao, and B. Liu, Journal of Physical Chemistry C 124, 4044 (2020).

\bibitem{PhysRevB.47.558} G. Kresse and J. Hafner, Phys. Rev. B 47, 558 (1993).
\bibitem{PhysRevB.54.11169}
G. Kresse and J. Furthmüller, Physical Review B 54, 11169 (1996).
\bibitem{PhysRevLett.77.3865}
J. P. Perdew, K. Burke, and M. Ernzerhof, Physical Review Letters 77, 3865 (1996).
\bibitem{PhysRevB.50.17953}
P. E. Bl¨ ochl, Physical Review B 50, 17953 (1994).
\bibitem{Togo2015}
A. Togo and I. Tanaka, Scripta Materialia 108, 1 (2015).
\bibitem{Martyna1992}
G. J. Martyna, M. L. Klein, and M. Tuckerman, The Journal of Chemical Physics 97, 2635 (1992).
\bibitem{Bykov2019}
M. Bykov, E. Bykova, G. Aprilis, K. Glazyrin, E. Koemets, I. Chuvashova, I. Kupenko, C. McCammon, M. Mezouar, V. Prakapenka, H. P. Liermann, F. Tasn{\'{a}}di, A. V. Ponomareva, I. A. Abrikosov, N. Dubrovinskaia, L. Dubrovinsky, Nature Communications 10, 2994 (2019)
\bibitem{Wilsdorf1948}
H. Wilsdorf, Acta Crystallographica 1, 115 (1948).
\bibitem{Crowhurst2006}
J. C. Crowhurst, Science 311, 1275 (2006).
\bibitem{Jia2021}
X. Jia, Q. Shao, Y. Xu, R. Li, K. Huang, Y. Guo, C. Qu, and E. Gao, npj Computational Materials 7, 211 (2021).

\end{thebibliography}
\end{document}